\documentclass[12pt,a4paper]{article}
\usepackage[english]{babel}
\usepackage[latin1]{inputenc}
\usepackage{amsmath}
\usepackage{amsfonts}
\usepackage{amssymb}
\usepackage{amsthm}

\begin{document}


\renewcommand{\vec}[1]{ \boldsymbol{#1} }

\title{Plasmons on adiabatically varying surfaces}
\author{Maria V Perel$^{1,2)}$ and Dmitry Yu Zaika$^{1)}$}

\maketitle
{{\footnotesize\center $^{1)}$ Department of Mathematical Physics, Physics Faculty, \\
St.Petersburg University, \\
 Ulyanovskaya 1-1, Petrodvorets, St.Petersburg, 198504, Russia \\
{\centerline {mailto: perel@mph.phys.spbu.ru, dmitry.zaika@gmail.com} }
{\footnotesize\center $^{2)}$ Ioffe Physico-Technical Institute, \\
\centerline{ Politekhnicheskaya 26, 194021 St. Petersburg, Russia }}}

\begin{abstract}
Surface plasmon polaritons (SSP),  moving along a smooth curved interface  between two isotropic media with slowly varying dielectric permittivities and magnetic permeabilities and supporting SSP, are studied theoretically.  Solutions of Maxwell equations are investigated within a small wavelength limit in the boundary layer of the  wavelength order near the surface. An explicit asymptotic formula for an EM wave  traveling along geodesics on the surface is obtained. An exponential factor reflects a distinction between the planar and curved cases. A curvature-dependent correction term in the exponent demonstrates a strong dependence on  transverse curvature and a weak dependence on  longitudinal curvature. The closer the parameters  to the resonance case, the more pronounced this tendency. It is found that the attenuation of the SPP in the case of  lossy media may be reduced by changing the curvature.

 If the signs of the magnetic permeability of the medium on both sides of the interface are opposite, the surface magnetic plasmon polariton may propagate.  Its short-wavelength asymptotics is found.
\end{abstract}


\newpage

Fabrication of nanoscale devices, applications to near-field optics, microscopy, and spectroscopy stimulated investigations of surface plasmon polariton propagation (see, for example \cite{Maier_2010, zayats2005nano, barnes2003surface, Raether-book}). Progress in metamaterials attracts attention to magnetic surface plasmons \cite{PhysRevB.71.195402}.
The aim of the present paper is to derive an asymptotic formula, which describes surface-curvature effects of the SPP.  Papers on the SPP on surfaces with  spherical and cylindrical geometries have appeared in the last century \cite{jensen1937eigenschwingungen}, \cite{miziumski1972utilization, ashley1974dispersion, Economou-Ngai, PhysRevE.50.4094}; now an interest in the dependence of properties of the SPP on the surface curvature  has been renewed.
  The propagation of the SPP on bends is studied in \cite{Hasegawa2004surface}, \cite{Hasegawa2007surface}.  Numerical methods of finding the SPP  on an arbitrary adiabatically varying surface,  using a curvilinear coordinate system, were developed in \cite{PhysRevA.80.043805} and \cite{valle2010geometric}.  The numerical  study of the influence of the curvature on the SPP characteristics was performed also in \cite{Liaw2008dispersion} and \cite{Vogel2010shape}.

  We give here an explicit asymptotic formula for the solution of the Maxwell equation near the interface between two media with smoothly varying dielectric and magnetic permittivities.
   We  use a curvilinear system  associated with geodesics, as was used in \cite{andronov2005asymptotic, Andronov2011effect}. Then we rewrite the Maxwell equation in  matrix form by analogy with  \cite{perel1990overexcitation} and apply the boundary layer method \cite{babich1979boundary}.
   The distinction from the planar case is in a slowly changing amplitude factor and a Berry phase, which provided a correction to the phase speed. In the case of lossy media, we obtain a curvature-dependent correction to the attenuation constant. The reduction of losses, as compared with the planar case, can be predicted with the formula. An explanation of this fact in the case of cylindrical bend has been done in \cite{Hasegawa2007surface}.

    A formula for a surface magnetic plasmon is found with the help  of the symmetry properties of Maxwell'a equations.

We solve the Maxwell equations for a monochromatic electric  ${\bf E}$ and magnetic   ${\bf H}$ fields of a frequency $\omega$
\begin{eqnarray*}
& {\rm rot\,}{\bf E} = ik \mu {\bf H}, \\
& {\rm rot\,}{\bf H} = -ik \varepsilon {\bf E}, \quad k = \omega/c,
\end{eqnarray*}
$c$ is the speed of light; the dielectric permittivity $ \varepsilon$ and the magnetic permeability   $\mu$   have a jump on the interface between two media and change slowly in each medium. We introduce  a curvilinear  coordinate system  $(s,a,n)$ associated with an interface, where $s$ is the length along the geodesic curve on the surface, the coordinate $a$ parameterizes distinct geodesics, the coordinate $n$ is measured along a normal at every point of the surface. We introduce a new  variable vector $\Psi = ( {\cal E},{\cal  H})^t,$ where ${\cal E}$ and ${\cal H}$ are covariant components of the electromagnetic field ${\cal E}= \hat{g} {\bf E},$  ${ \cal H}= \hat{g}{\bf H},$  $\hat{g}$ is the metric tensor.  In terms of parameters of the surface, namely, the longitudinal radius of curvature $\rho,$ the radius of curvature in transverse direction $\rho_t,$
the divergence of geodesics on the surface $h,$ and the torsion $\tau,$ the metric tensor is expressed as follows:
\begin{align*}
\hat{g} =
  \begin{pmatrix}
    \big(1+\frac{n}{\rho}\big)^2+\tau^2n^2  & -h\tau \big(2n+\frac{n^2}{\rho^2}+\frac{n^2}{\rho_t^2}\big) & 0\\
    -h\tau \big(2n+\frac{n^2}{\rho^2}+\frac{n^2}{\rho_t^2}\big) & h^2\big((1+\frac{n}{\rho_t})^2+\tau^2n^2\big)
     & 0\\     0 & 0 & 1\\
  \end{pmatrix}.
\end{align*}
We rewrite the Maxwell equations in  matrix form by analogy with  \cite{Felsen-book},\cite{perel1990overexcitation}
\begin{eqnarray}
\label{maxwell_n}
\hat{P}{\bf\Psi} + i\hat{\Gamma}_n\partial_n {\bf\Psi} = -i \left(\hat{\Gamma}_s\partial_s - \hat{\Gamma}_a\partial_a \right){\bf\Psi},
\end{eqnarray}
where
\begin{align*}
&\,\hat{P}=\left(\begin{matrix}
                           \hat{\bf \alpha}{ \varepsilon} &  \hat{ 0} \\
                           \hat{ 0} & \hat{\bf \alpha} {\mu}\\
                           \end{matrix}\right),\quad
                         \hat{\Gamma}_n =
                         \left(\begin{matrix}
                           \hat{0} & \hat{\gamma}_n \\[2pt]
                            -\hat{\gamma}_n & \hat{0}\\[2pt]
                          \end{matrix}\right),\\
                        &  \hat{\Gamma}_a =
                         \left(\begin{matrix}
                           \hat{0} & \hat{\gamma}_a \\[2pt]
                            -\hat{\gamma}_a & \hat{0}\\[2pt]
                          \end{matrix}\right),\quad
                          \hat{\Gamma}_s =
                         \left(\begin{matrix}
                           \hat{0} & \hat{\gamma}_s \\[2pt]
                            -\hat{\gamma}_s & \hat{0}\\[2pt]
                          \end{matrix}\right).
  \end{align*}
  Here $\hat{\alpha}$ is defined as  $\hat{\alpha} = ({\rm det} \hat{g})^{1/2} \hat{g}^{-1}$,
\begin{align*}
  \varepsilon = \begin{cases}
        \varepsilon_+,  \quad n > 0 \\[5pt]
        \varepsilon_-,  \quad n < 0 \\[5pt]
        \end{cases}, \quad
     \mu = \begin{cases}
        \mu_+,  \quad n > 0 \\[5pt]
        \mu_-,  \quad n < 0 \\[5pt]
        \end{cases}, \quad
\end{align*}
 \begin{align*}
 \hat{\gamma}_n =      \left(\begin{matrix}
                           0 & 1 & 0 \\[2pt]
                           -1 & 0& 0 \\[2pt]
                             0& 0& 0 \\[2pt]
                          \end{matrix}\right),
                          \hat{\gamma}_s =      \left(\begin{matrix}
                           0 & 0 & 0 \\[2pt]
                           0 & 0& 1 \\[2pt]
                           0 &-1& 0 \\[2pt]
                          \end{matrix}\right),
                          \hat{\gamma}_a =      \left(\begin{matrix}
                           0 & 0 & 1 \\[2pt]
                           0 & 0& 0 \\[2pt]
                           -1 & 0& 0 \\[2pt]
                          \end{matrix}\right).
 \end{align*}

The investigation of the exactly solvable problem of diffraction on a cylinder shows that the plasmon wave is concentrated in the boundary layer of the wavelength order near the surface.
We make  an assumption $n=O(\frac{1}{k}),$ where $k$ is the wavelength in vacuum. We suppose that the unit of the length is the typical radius of curvature $\rho$ and
 $k  \gg 1.$  We decompose the matrix $\hat{\alpha}$ in  $k^{-1}$ and obtain
 \begin{equation}
 \hat{P} = \hat{P}^{(0)} + \frac{1}{k}\hat{P}^{(1)} + O\left(\frac{1}{k^2}\right),
 \end{equation}
\begin{align*}
  \hat{P}^{(0)} =   \left(\begin{matrix}
                            { \varepsilon} { \hat{H}} &  \hat{0} \\
                           \hat{0} & {\mu} {\hat{H}} \\
                           \end{matrix}\right),\qquad
 \hat{P}^{(1)} = \left(\begin{matrix}
                          {\varepsilon}{ \hat{M}} & \hat{0} \\[2pt]
                           \hat{0} &  {\mu}{\hat{M}} \\[2pt]
                          \end{matrix}\right),
 \end{align*}
\begin{align*}
{\hat{M}} = \ell \left(\begin{matrix}
                           -h(\frac{1}{\rho} - \frac{1}{\rho_t}) & 2 \tau & 0 \\
                           2 \tau &  h^{-1}(\frac{1}{\rho} - \frac{1}{\rho_t}) & 0 \\
                           0 & 0& h(\frac{1}{\rho} + \frac{1}{\rho_t})  \\
                          \end{matrix}\right), \qquad
\end{align*}
${ \hat{H}} = {\rm diag}(h, h^{-1}, h)$.
We seek the solution of (\ref{maxwell_n}) in the form
\begin{align}
\label{sol-form}
& {\bf \Psi} = e^{i k \int\limits_0^s\beta(s'\!,\, a)\,ds'}{\bf\Phi}(s,a,\ell), \quad
{\bf\Phi}=\sum\limits_{j=0}^{\infty} k^{-j} {\bf\Phi}^{(j)}.
\end{align}
Substituting (\ref{sol-form}) into the Maxwell equations and equating terms of equal orders of $k$,   we obtain
\begin{align}
\left( \hat{P}^{(0)} + i{\hat{\Gamma}_n}\partial_n \right) {\bf\Phi}^{(0)} & = \beta \hat{\Gamma}_s {\bf\Phi}^{(0)}, \label{mm1} \\
\left( {\hat{P}^{(0)}} + i{\hat{\Gamma}_n}\partial_n \right) {\bf\Phi}^{(1)} & = \beta {\hat{\Gamma}_s} {\bf\Phi}^{(1)}
- {\hat{P}^{(1)}} {\bf\Phi}^{(0)} \nonumber\\
& \quad  - i \left(\hat{\Gamma}_s\partial_s - \hat{\Gamma_a}\partial_a \right){\bf\Phi}^{(0)}. \label{mm2}
\end{align}
The systems of equations split into two systems in the upper and lower media. The condition of the continuity of the tangential components of the electrical and magnetic fields on the interface between the media yields a relation between the coefficients of the solutions. After simplifications we obtain
 \begin{align} \label{zero-order}
   & {\bf\Phi}^{(0)}  = C(s,a){\boldsymbol {\varphi}}, \quad\\ & {\boldsymbol{\varphi}}  = \begin{cases}
    e^{-\kappa_+\ell} \cdot\left(
    \frac{i\kappa_+}{h \varepsilon_+}\,,0\,,-\frac{\beta}{h \varepsilon_+}\,,0\,,1\,,0\right)^{\mathrm{t}}, \quad l>0, \\
    e^{ \kappa_-\ell} \cdot\left(
    - \frac{i\kappa_-}{h \varepsilon_-}\,,0\,,-\frac{\beta}{h \varepsilon_-}\,,0\,,1\,,0\right)^{\mathrm{t}}, \quad l<0,
\end{cases} \label{field-zero}
\end{align}
$
 \kappa_{\pm}=  \sqrt{\beta^2-\epsilon_{\pm}\mu_{\pm}},
$
and the square root  with a positive real part is chosen.
The condition of continuity of the first component of ${\bf\Phi}^{(0)}$ yields the relation
$
     {\kappa_{+}}/{\epsilon_+} = - {\kappa_{-}}/{\epsilon_-},
$
which can be satisfied if $\varepsilon_1$ and $\varepsilon_2$ have opposite signs. This condition is equivalent to the requirement
\begin{align} \label{SP-b}
   \beta^2= \frac{( \mu_+ \varepsilon_- - \mu_-\varepsilon_+) \varepsilon_+ \varepsilon_- }{\varepsilon_-^2 - \varepsilon_+^2},
\end{align}
where $\beta$ coincides with the propagation constant of a plasmon on the plane interface between two homogeneous media.
 When the moduli of $\varepsilon_1$ and $\varepsilon_2$ coincide, a plasmon resonance occurs, $\beta$ and $\kappa_{\pm}$ tend to infinity, and the field may be concentrated in a subwavelength layer near the interface.

 The function $C(s,a)$ of the coordinates on the surface should be defined from the equation (\ref{mm2}).
We note that if $\varepsilon$ and $\mu$ are real-valued, then the operators on the left-hand side and right-hand side of (\ref{maxwell_n}) are Hermitian if the scalar product
$({\bf f},{\bf g})$ is defined by the standard way
\begin{align*}
({\bf f},{\bf g}) = \int\limits_{-\infty}^{\infty} \sum \limits_{k=1}^{6} f^*_j(l)g_j(l) d l;
\end{align*}
here, the superscript * stands for complex conjugation.
 The solvability condition of (\ref{mm2}) has the  form
 \begin{align}
i\left( {\bf\Phi}^{(0)}\,,\hat{\Gamma}_s\partial_s{\bf\Phi}^{(0)} \right)
 + i \left( {\bf\Phi}^{(0)}\,,\hat{\Gamma}_a\partial_a{\bf\Phi}^{(0)} \right) + \quad \nonumber\\ ({\bf\Phi}^{(0)},\hat{P}^{(1)}{\bf\Phi}^{(0)})  = 0.
\label{solvab}
\end{align}
We note that $\left( {\bf\Phi}^{(0)}\,,\hat{\Gamma}_s {\bf\Phi}^{(0)} \right)$ and $\left( {\bf\Phi}^{(0)}\,,\hat{\Gamma}_a {\bf\Phi}^{(0)} \right)$ are average fluxes of energy through areas of infinite height and unit width, which are orthogonal to $s$ and $a$ directions, respectively.
 Upon substitution of (\ref{zero-order}) into this equation, we obtain an expression for $C(s,a)$:
 \begin{equation}\label{c--}
 C(s,a) = D \left( {\boldsymbol{ \varphi}}\,,\hat{\Gamma}_s{\boldsymbol{\varphi}} \right)^{-1/2}
 \exp {\left(i \int\limits_{s_0}^{s} \frac{({\boldsymbol{\varphi}},\hat{ P}^{(1)}{\boldsymbol{\varphi}})}{( {\boldsymbol{\varphi}}\,,\hat{ \Gamma}_s{\boldsymbol{\varphi}} )} \, d\tilde{s} \right)},
 \end{equation}
 where $D$ may depend only on $a$.
 After calculations we obtain
 \begin{align}
 & \left( {\boldsymbol{\varphi}}\,,\hat{\Gamma}_s{\boldsymbol{\varphi}} \right) = \frac{\beta\varepsilon_+}{h \kappa_+}\left(\frac{1}{\varepsilon_+^2} - \frac{1}{\varepsilon_-^2} \right),  \label{h1}\\
 & \left({\boldsymbol{\varphi}}\,,\hat{P}^{(0)}{\boldsymbol{\varphi}}  \right) =  \frac{1}{2 h \rho_t}\left(\frac{1}{\varepsilon_+} - \frac{1}{\varepsilon_-}\right)+
 \frac{1}{2h\rho}\left( \frac{\mu_+}{\kappa_+^2} - \frac{\mu_-}{\kappa_-^2}\right). \label{h2}
  \end{align}
 Gathering formulas (\ref{sol-form}), (\ref{zero-order}), (\ref{c--}), (\ref{h1}), (\ref{h2}), we obtain the principal term of the asymptotics
\begin{align}
 {\bf \Psi}^{(0)} = & {\bf A}\ \exp{\left(i k \int\limits_0^s \beta\,ds'\right)}\exp{ \left( i \int\limits_{s_0}^s
 (\frac{\sigma}{\rho} + \frac{\sigma_t}{\rho_t}  )\, ds \right)} \begin{cases}  &\exp{(- \kappa_+ \ell)}  ,\quad \ell>0,\\ &\exp{( \kappa_- \ell)}, \quad \ell<0,\end{cases}
  \label{res-C}  \\
  {\bf A} = & \qquad \left(\frac{i}{h^{1/2}}\frac{(\mu_+\varepsilon_+ - \mu_-\varepsilon_-)^{3/4}(-\varepsilon_+\varepsilon_-)^{3/4}}{(\mu_-\varepsilon_+ - \mu_+\varepsilon_-)^{1/4}(\varepsilon_-^2 - \varepsilon_+^2)}, \qquad \qquad \qquad  0, \quad \right.\nonumber\\
 &\left. - \frac{(\mu_+\varepsilon_+ - \mu_-\varepsilon_-)^{1/4}(\mu_-\varepsilon_+ - \mu_+\varepsilon_-)^{1/4}
 (-\varepsilon_+\varepsilon_-)^{5/4} }   {h^{1/2}\varepsilon_{\pm}(\varepsilon_-^2 - \varepsilon_+^2)},\qquad  \right. 0,  \nonumber\\
  & \left. \quad h^{1/2} \left(\quad \frac {(\mu_+\varepsilon_+ - \mu_-\varepsilon_-) \quad (-\varepsilon_+\varepsilon_-)^3} {(\mu_-\varepsilon_+ - \mu_+\varepsilon_-)\quad (\varepsilon_-^2 - \varepsilon_+^2)^2} \quad \right)^{1/4}, \qquad   \quad 0     \right), \label{A}\\
 \sigma  = & \qquad\frac{(\mu_+ \varepsilon_-^2 - \mu_-\varepsilon_+^2)}{2 \left( \quad \varepsilon_+(-\varepsilon_-)
  ({\mu_-}{\varepsilon_+} - {\mu_+}{\varepsilon_-}) \,(\mu_+ \varepsilon_+ - \mu_-\varepsilon_-) \quad \right)^{1/2}}, \quad \label{res-sig}\\
 \sigma_t  = & \qquad \frac{1}{2}\left( \frac{\mu_+ \varepsilon_+ - \mu_-\varepsilon_-}{\mu_-\varepsilon_+ - \mu_+\varepsilon_-}\right)^{1/2} \frac { (-\varepsilon_-\varepsilon_+)^{1/2}} {(-\varepsilon_+ - \varepsilon_-)}.
  \label{res-sigma}
\end{align}
Here, the $\varepsilon_{\pm}$ stand for $\varepsilon_+$ if $\ell>0$ and for $\varepsilon_-$ if $\ell<0.$
We has written (\ref{solvab}) assuming for simplicity that $\varepsilon$ and $\mu$ are real-valued. In lossy media operators in (\ref{maxwell_n}) are non-Hermitian. The solvability condition of (\ref{mm1}) differs from (\ref{solvab}). In all scalar products, the function ${\bf \Phi^{(0)}}$ standing on the left must be replaced by a function ${\bf \Phi^{(0)+}},$ which solves an equation adjoint to (\ref{mm2}). Analogously, the function ${\boldsymbol{\varphi}}$ standing on the left in scalar products in (\ref{c--}) must be replaced by ${\boldsymbol{\varphi^+}}.$ To find ${\bf \Phi^{(0)+}}$ we should take ${\bf \Phi^{(0)}}$ corresponding to $\beta^*,$  $\varepsilon^*,$ and $\mu^*.$  Upon calculations, we obtain the same result (\ref{res-C}), (\ref{A}),  (\ref{res-sig}), (\ref{res-sigma}).

The formulas can be simplified in the  case $\mu_+ = \mu_- = 1;$  then formulas (\ref{res-sigma}) read
 \begin{equation}\label{mu-1}
 \sigma = - \frac{\varepsilon_+ + \varepsilon_-}{\sqrt{\varepsilon_+(-\varepsilon_-)}},  \qquad
 \sigma_t =  \sigma^{-1}.
 \end{equation}

The amplitude factor $A$ varies adiabatically, because  $h$, $\varepsilon,$ and $\mu$ may vary. The first exponential term does not feel the curvature, it is the same as in the case of homogeneous media. The second exponential term is a correction term, which is curvature dependent.
 It is real-valued in the case of nonabsorbing medium, and it represents the Berry phase.
Both formulas (\ref{res-sigma}) and (\ref{mu-1}) demonstrate a  strong dependence on the transverse curvature radius near the plasmon resonance $\varepsilon_+ + \varepsilon_- =0$. Moreover, the correction has a stronger  singularity than the propagation constant $\beta$
 near the plasmon resonance.   The influence of the longitudinal radius $\rho$ tends to zero, when the conditions are close to the plasmon resonance and $\mu_+=\mu_- >0$ according to (\ref{mu-1}). If $\mu_+$ and $\mu_-$ differ and are positive, the influence of $\rho$ may be important but does not have a resonance character.

A surprising result is the possibility of reducing losses by changing the curvature in the case where we deal with lossy media. The correction terms become complex-valued. They contribute to the attenuation of the SPP. The curvature radii have different signs for concave and convex surfaces. Therefore the correction terms may give not only attenuation but an increase  in the wave as well. This effect is in agreement with  numerical results on the circular bend \cite{Hasegawa2004surface}, \cite{Hasegawa2007surface}.  An explanation of this phenomenon in the case of curved boundary the metal-dielectric is given in \cite{Hasegawa2007surface}. There are two loss mechanisms: the bending loss through the radiation into the dielectric and absorption loss in the metal. The bend increases the loss because of the radiation into the dielectric and decrease it because the bend pushes the wave out of the metal where the wave travels with attenuation. These effects can be comparable and can result in reduction of losses comparing with a planar surface \cite{Hasegawa2007surface}.

Qualitatively new results can be obtained in metamaterials where magnetic permeabilities may have opposite signs.
 Then the surface magnetic plasmon (SMP) may exist. To obtain a formula for the  surface magnetic plasmon, we should use the symmetry of Maxwell equations and interchange $\varepsilon$ and $\mu$, ${\bf E}$ and $-{\bf H}$ in the result (\ref{c--})-(\ref{res-sigma}). We use the notation $\beta_m,$ $\sigma_m$ and $\sigma_{tm}$ for SMP instead of the notation $\beta,$ $\sigma,$ and $\sigma_t$ for SSP. The propagation constant  $\beta_{m}$ reads
\begin{align} \label{SMP-b}
   \beta_m^2 = \frac{( \mu_- \varepsilon_+ - \mu_+\varepsilon_-) \mu_+ \mu_- }{\mu_-^2 - \mu_+^2}.
\end{align}
The dependence of the exponent in the correction term on $\rho$ and $\rho_t$ is controlled by the parameters
\begin{align*}
&\sigma_m  = \frac{(\varepsilon_+ \mu_-^2 - \varepsilon_-\mu_+^2)}{2 \left(  \mu_+(-\mu_-)
  ({\varepsilon_-}{\mu_+} - {\varepsilon_+}{\mu_-}) \,(\varepsilon_+ \mu_+ - \varepsilon_-\mu_-) \, \right)^{1/2}}, \\
 &\sigma_{tm}  = \frac{1}{2}\left( \frac{\varepsilon_+ \mu_+ - \varepsilon_-\mu_-}{\varepsilon_-\mu_+ - \varepsilon_+\mu_-}\right)^{1/2} \frac { (-\mu_-\mu_+)^{1/2}} {(-\mu_+ - \mu_-)}.
  \end{align*}

If  $\varepsilon_-$ and $\mu_-$ are negative, both types of plasmons may exist.
The special cases where  $\mu_- \varepsilon_+ = \mu_+\varepsilon_-$ or $\mu_+ \varepsilon_+ = \mu_-\varepsilon_-$  correspond to the case of the coincidence of propagation constants (\ref{SP-b}) and (\ref{SMP-b}). In this case, both asymptotic formulas are not applicable, plasmons do not propagate independently,  and new formulas describing the process of plasmon interactions should be obtained.  If parameters are close to degeneration but the formulas are still applicable,  the correction terms may be very significant.

 The  authors thank A.M. Monakhov, V.N. Trukhin,  D. Bouche, and  I. V. Andronov for valuable discussions. The second author would also like to convey thanks to the ENS Cachan for providing the financial means and research facilities.

 \bibliographystyle{unsrt}
\bibliography{Lit}

 \end{document}